\documentclass[
    ,final            
  ]
  {aipproc}

\layoutstyle{6x9}
\newcommand{\la}{\lambda}
\newcommand{\be}{\beta}
\newcommand{\ka}{\kappa}
\newcommand{\beano}{\begin{eqnarray*}}
\newcommand{\eeano}{\end{eqnarray*}}

\begin{document}

\begin{flushright}
UdeM-GPP-TH-02-101
\end{flushright}

\title{Magnetic behaviour of SO(5) superconductors}

\author{R. MacKenzie}{
  address={Laboratoire-Ren\'e-J.-A.-L\'evesque, Universit\'e de Montr\'eal\\
           C.P.~6128, Succ.~Centre-ville, Montr\'eal, Qc H3C 3J7, Canada}
}

\begin{abstract}
The distinction between type I and type II superconductivity is
re-examined in the context of the SO(5) model recently put forth by
Zhang. Whereas in conventional superconductivity only one parameter
(the Ginz\-burg-Landau parameter $\kappa$)
characterizes the model, in the SO(5) model there are two essential
parameters. These can be chosen to be $\kappa$ and another parameter,
$\beta$, related to the doping.
There is a more complicated relation between
$\kappa$ and the behaviour of a
superconductor in a magnetic field. In particular, one can find type I 
superconductivity even when $\kappa$ is large, for appropriate values 
of $\beta$.
\end{abstract}

\maketitle


\section{Introduction}

In this talk, recent work on magnetic properties of the SO(5) model
  of high-temperature superconductivity (HTSC) will be presented. After
  reviewing the case of
  conventional superconductivity and some relevant facts of HTSC, the
  SO(5) model will be introduced. The behaviour of a system
  described by this model when placed in a magnetic field will be
  analysed. The main conclusion is that in strongly underdoped
  superconductors, the critical value of the Ginzburg-Landau (GL)
  parameter $\kappa$ can be much larger than the conventional
  value. Thus, a large value of $\kappa$ can be associated with a type
  I superconductor. The application of these ideas to HTSC will be
  discussed briefly. This work forms the bulk of Refs. \cite{us1,us2}.

\section{Preliminaries}

\subsection{Conventional superconductivity}

In this section we briefly review some features of conventional
superconductivity (SC). This material is well-known, and can be found in
greater detail in almost any introductory
SC textbook, such as
Tinkham \cite{tinkham}.

The phase transition in
a conventional superconductor can be described at low energies by an
effective theory, known as a Ginzburg-Landau (GL) theory, written in
terms of the SC order parameter $\phi$ (a complex field
representing the Cooper pair amplitude) and the electromagnetic
field. This GL theory can be expressed in terms of a Helmholtz free
energy, which takes the following form:
\[
{F}=\int{d\mathbf{x}} \left\{ f_n - \frac{a^2_1}{2} |\phi|^2 + \frac{b}
{4} |\phi|^4 + \frac{1}{2m^*} \left|\left(-i\hbar\nabla - \frac{e^*\mathbf{A}}
{c}\right)\phi\right|^2+\frac{\mathbf{h}^2}{8\pi}\right\}.
\]
Here, $f_n$ is a constant, ${\mathbf{h}}= \nabla \times \mathbf{A}$ is 
the microscopic magnetic field and $a_1$, $b$ are parameters. The minimum 
of the potential is $|\phi|^2 = a^2_1/b \equiv v^2$.

There are two characteristic length scales in this model: the
coherence length $\xi= (\hbar ^2 / m^*a^2_1)^{1/2}$ and the magnetic
field penetration depth $\la=(m^* c^2 / 4\pi e^{*^{2}}
v^2)^{1/2}$. These are, roughly, the Compton wavelengths of the scalar
and electromagnetic fields, respectively. By scaling out all
dimensionful quantities, one finds that the behaviour of a
SC described by the above free energy is determined by one
dimensionless parameter, the GL parameter $\kappa=\lambda/\xi$.
Some typical values for this parameter appear in Table \ref{params}.
\vspace{8pt}
  \begin{table}[htb]
  \begin{tabular}{rccc}
   \hline
   &\tablehead {1}{c}{b}{$\la$ (\r{A})}
   &\tablehead {1}{c}{b}{$\xi$ (\r{A})}
   &\tablehead {1}{c}{b}{$\kappa$}\\
   \hline
   Simple metal&300&1000&.3\\
   Alloy&2000&50&40\\
   High-T$_{\rm c}$&2000&20&100\\
   \hline
  \end{tabular}
  \caption{Typical parameter values for various categories of
  superconductors.}
  \label{params}
  \end{table}
\vspace{6pt}

There are two very different classes of (conventional) SCs, depending on
the value of $\kappa$. If $\kappa<1/\sqrt2$, the material is said to
be type I, while if $\kappa>1/\sqrt2$, it is said to
be type II. 
The behaviour when a magnetic field is applied to a
superconductor differs greatly for these two classes. 

\begin{figure}[h]
  \includegraphics[angle=270,width=.6\textwidth]{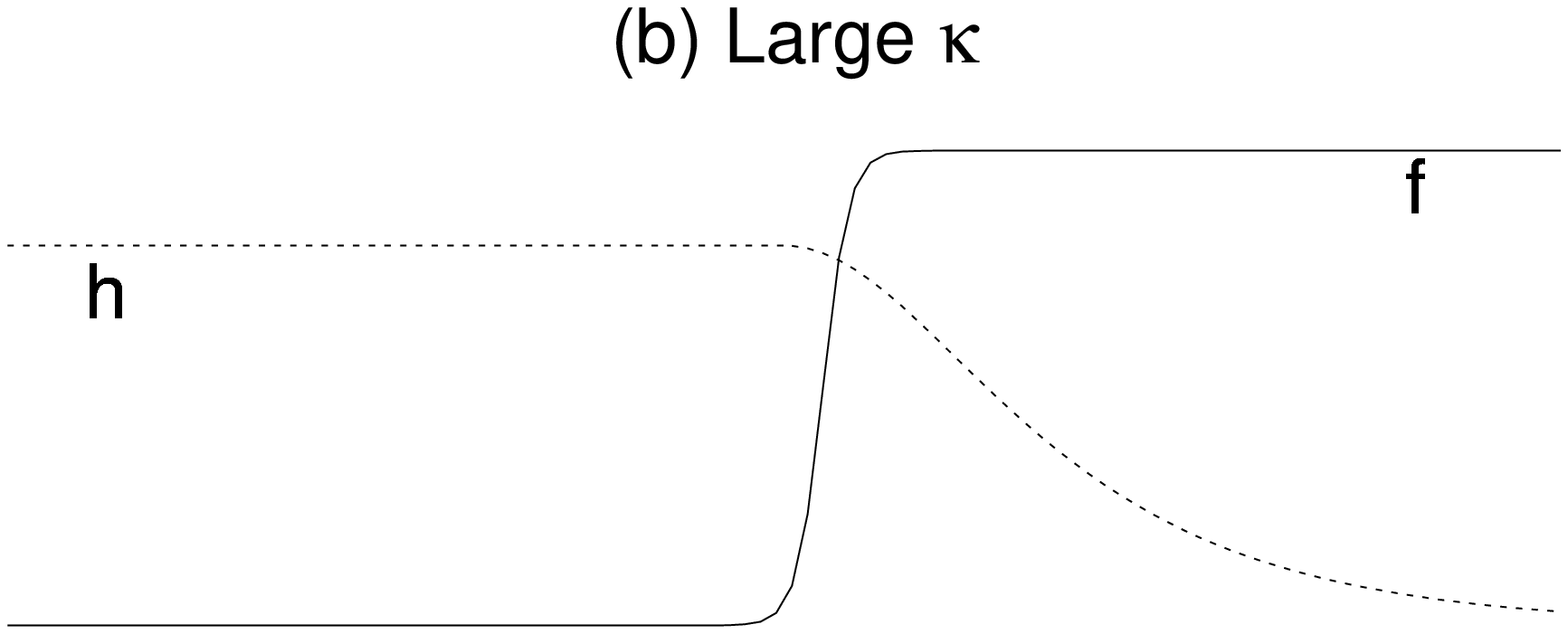}
  \caption{Field profiles at SC/normal surface for (a) Type I and (b)
  Type II superconductors. $f$ is a rescaled $\phi$. The slow
  variation of $f$ when $\kappa\ll1$ gives rise to a positive
  contribution to the surface energy; the slow variation of $h$ has
  the opposite effect in the opposite limit.
(Figure from "Magnetic Properties of SO(5) Superconductivity" by
  M. Juneau, R. MacKenzie and M.-A. Vachon, in Annals of Physics,
  Volume 298, 421, copyright 2002, Elsevier Science (USA), reproduced
  by permission of the publisher.)
}
\label{surface}
\end{figure}

There are several ways to see this. One way is to consider a
configuration where a magnetic field equal to the so-called
thermodynamic critical field $H_{c}^0$ is applied to the
superconductor.\footnote{If a weak magnetic field is applied to a
superconductor, the field is expelled; if a strong field is applied,
SC is destroyed and the field penetrates the
material. The critical field is the transitional value, i.e., that
where the (Gibbs) free energy of the normal phase in the magnetic
field is equal to that of the SC phase in the field's absence.}
The sign of the
energy of a surface separating SC and normal regions
is the telling quantity. If it is positive, the system
would prefer to minimize the amount of surface for a given magnetic
flux; this is achieved if the flux penetrates in a macroscopic
region. In contrast, if the surface energy is negative, the flux will
form a lattice of flux tubes of the minimum allowable flux. One can
calculate numerically the surface energy of a boundary between SC and
normal regions. Even quantitatively, one can argue that for small
$\kappa$ the surface energy is positive, while for large $\kappa$ it
is negative (see Figure \ref{surface}).

\begin{figure}[htb]
  \includegraphics[angle=270,width=.6\textwidth]{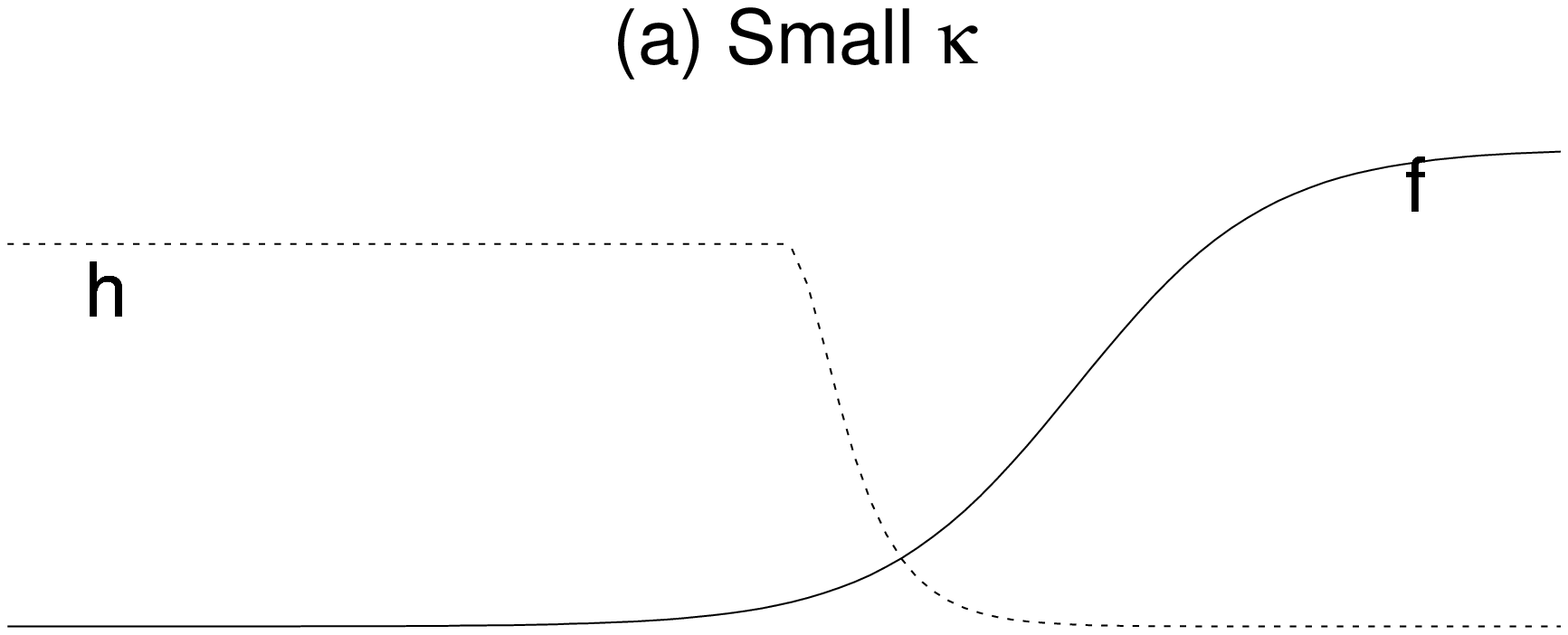}
\end{figure}

An alternative way to determine the distinction between type I and
type II superconductors is to consider the energetics of vortices of
varying winding number. Since the magnetic flux of a vortex is
proportional to its winding number, the energy per unit winding number
indicates whether it is energetically favourable for a given amount of
flux to penetrate in many unit-winding-number vortices or in one
large vortex. The former will be the case if the energy per unit
winding number is of positive slope, while the latter will be true if
it is of negative slope (see \cite{us1}).

\begin{figure}[htb]
  \includegraphics[height=.3\textheight]{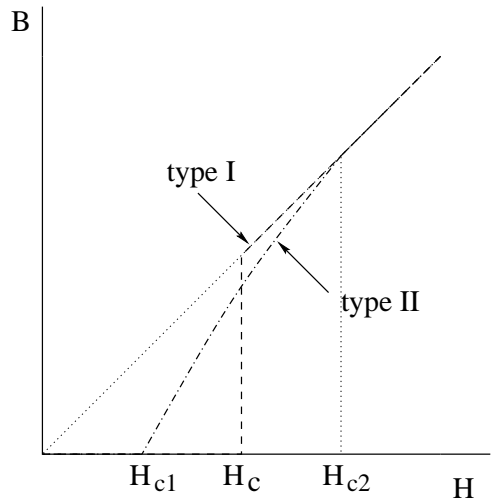}
  \caption{$B$ vs. $H$ for type I and type II superconductors.}
\label{mag}
\end{figure}

The magnetization curves of type I and type II superconductors
also highlight their different behaviour in a magnetic field (see
Figure \ref{mag}). As the magnetic field is increased in a type I
superconductor, the field is completely expelled until $H_{c}^0$ is
reached, at which point SC is destroyed
macroscopically. In contrast, in a type II superconductor, at a lower
field $H_{c1}^0$ the field starts to penetrate the superconductor in
vortices; when the upper critical field $H_{c2}^0$ is reached,
SC is finally destroyed. These critical fields vary as
a function of $\kappa$; one finds $H_{c2}^0=\sqrt{2}\kappa
H_{c}^0$. The transition point between type I and type II
SC occurs when these two critical fields are equal,
which occurs at $\kappa=\kappa_c=1/\sqrt2$.
\vspace{12pt}

\subsection{High-temperature superconductivity}

\begin{figure}[htb]
  \includegraphics[height=.25\textheight]{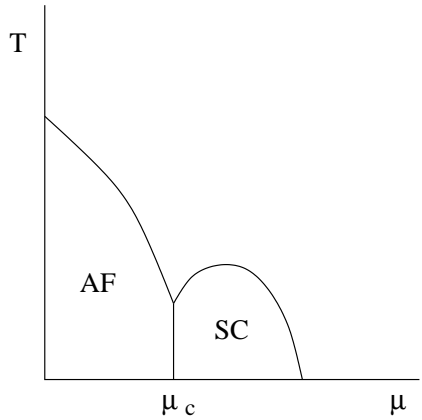}
  \caption{Approximate phase diagram for high-temperature
  superconductors and for SO(5) superconductivity.}
\label{phasediag}
\end{figure}

In this section, a couple of relevant facts of HTSC are presented.
The key observation which leads to the SO(5) model of HTSC is that
these materials exhibit two very different phases at low temperature,
depending on the degree of doping. At sufficiently high doping, one
sees SC, while at lower values of the doping (including
the undoped case)
the materials are antiferromagnetic (AF), as
shown in Figure \ref{phasediag}.

\vspace{12pt}

An important property of HTSCs is that, as mentioned
above (see Table \ref{params}), they
are ``highly type II'', i.e., $\kappa$ is very much larger
than its critical value. (Values in the vicinity of 100 are typical;
the lowest value we have seen reported is 17.)

However, as we shall see presently, in the SO(5) model, $\kappa$ alone
does not
determine the magnetic behaviour (i.e., the type) of a
superconductor. Indeed, it is possible that, in spite of having a very
large $\kappa$, HTSCs (if described by the SO(5) model)
might exhibit type I behaviour under certain conditions.

\section{SO(5) Superconductivity}

\subsection{Motivation; Ginzburg-Landau model}

As mentioned above, the presence of both AF and
SC in HTSCs suggests the possibility of a sort of
unification of these phenomena, both of which involve spontaneous
symmetry breaking. This possibility was put forth by Zhang
\cite{zhang}, who wrote down a model in terms of a five-component real
order parameter. The five components are the real and imaginary
components of the SC order parameter $\phi$ and the three components
of the AF order parameter $\eta$.
A GL theory which has an approximate SO(5) rotational symmetry can
then be written down; the free energy is
\[
 {F}=\int{d\mathbf{x}} \Bigg\{ 
  \frac{\mathbf{ h}^2}{8\pi}+
  \frac{\hbar^2}{2m^*} \left|\left(\nabla
  + \frac{ie^*\mathbf{A}}{\hbar c}\right)\phi\right|^2
 +\frac{\hbar^2}{2m^*} \left(\nabla\eta\right)^2
 +V(\phi,\eta)\Bigg\},
\]
where the potential is
\[
V(\phi,\eta)=-{{a_1}^2\over2}\phi^2-{{a_2}^2\over2}\eta^2
+{b\over4}(\phi^2+\eta^2)^2.
\]
There are now three relevant length scales: $\la$, $\xi$ and $\xi'$
(the characteristic length of the $\eta$ field). Rescaling now reduces
the number of essential parameters to two, which can be taken to be
$\kappa$ and $\be\equiv({a_2}^2/{a_1}^2)$. The latter is related to
the doping; $\be=1$ at the AF-SC boundary, and $\be<1$ in the SC phase.

\begin{figure}[htb]
  \includegraphics[height=.35\textheight]{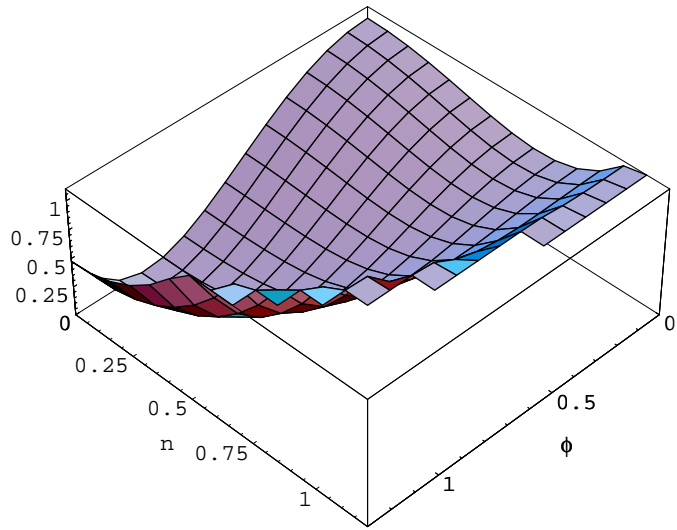}
  \caption{SO(5) potential as a function of $\phi$ and $\eta$.}
\label{potl}
\end{figure}

The potential for $\be<1$
is shown in Figure \ref{potl}. There are two important features of
the potential. First, it is minimized at a nonzero value of $\phi$
and $\eta=0$, so the ground state is indeed SC. Second,
if $\phi$ is somehow forced to be zero, then $\eta$ will be nonzero.
\vspace{12pt}

In fact, there are two situations when $\phi$ is indeed forced to be
zero: firstly, in the core of a vortex \cite{zhang,abkz,abbg,us1}, and
secondly, if the
superconductor is placed in a
sufficiently strong magnetic field.

\subsection{Magnetic properties}

This ``induced antiferromagnetism'' can have a dramatic effect on the
critical fields, and on the type of superconductor described by the
model. The effect on the
thermodynamic critical field $H_c$ arises because it
is found by comparing the Gibbs free energies of SC and
AF (rather than normal) states. The other critical
fields $H_{c1,2}$ are affected because vortex energetics are affected
by the AF core.

\begin{figure}[htb]
  \includegraphics[width=.6\textwidth]{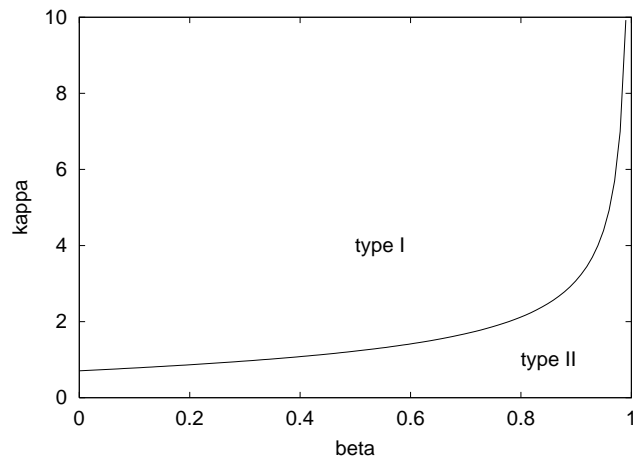}
  \caption{Curve delineating type I vs type II superconductivity in
  the $\kappa$-$\beta$ plane.
(Figure from "Magnetic Properties of SO(5) Superconductivity" by
  M. Juneau, R. MacKenzie and M.-A. Vachon, in Annals of Physics,
  Volume 298, 421, copyright 2002, Elsevier Science (USA), reproduced
  by permission of the publisher.)
}
\label{bkplane}
\end{figure}

Both $H_c$ and $H_{c2}$ can be calculated analytically \cite{us2}:
\beano
 H_c=H_c^0\sqrt{1-\be^2};\\
 H_{c2}=H_{c2}^0(1-\be)=\sqrt2\kappa H_c^0(1-\be).
\eeano
As in the conventional case, equality of these critical fields
indicates the boundary of type I/II behaviour. We can thus obtain a
curve in the $\be$-$\ka$ plane which represents the boundary
separating type I and type II behaviour (Figure
\ref{bkplane}). Analytically, this curve is given by
\[
\kappa_c(\beta)
 = {1\over\sqrt{2}}\sqrt{{1 + \beta\over 1- \beta}}.
\]
This expression is confirmed (numerically) by analysis of surface
energy and vortex energetics.

\subsection{Application to high-temperature superconductivity}

In HTSC, as mentioned above, $\kappa$ is typically of order 100; thus,
HTSCs are considered highly type II. However, as the previous section
demonstrates, in the SO(5) model $\beta$ also plays a role in the
nature of the superconductor. In particular, for any
$\kappa>1/\sqrt2$, if $\beta$ is sufficiently large, the material is
type I. We can invert the above expression for $\kappa_c(\beta)$ to
obtain the following expression for $\beta_c(\kappa)$, valid if
$\kappa\gg1$ $\beta_c(\kappa)\simeq 1-\kappa^{-2}$.
For example, if $\kappa=100$, the material is a type I superconductor
if $\beta<0.9999$ while it is type II if $\beta>0.9999$. (Since
$\beta<1$ in the SC state, there is only a minute range of $\beta$
corresponding to type II.) If $\kappa=17$ (the smallest value reported
for a HTSC), $\beta>0.996$ for the material to be type II -- still a
small window, but much greater than that for $\kappa=100$.

The parameter $\beta$ is related to measurable quantities:
$
\beta=1-8m^*\hbar^{-2}\xi(\mu^2-\mu_c^2),
$
where $\mu$, $\mu_c$ are the chemical potential and the critical
chemical potential (that at the SC-AF boundary) and $\xi$ is the
charge susceptibility.

Thus, we have the possibility of
a fairly dramatic test of the SO(5) model; however, the
experimental situation in the underdoped region
appears rather delicate. For example,
the appearance of inhomogeneities (stripe formation, phase separation)
could mask the appearance of type I behaviour. Nonetheless, since the
degree to which a superconductor is type II is reduced as the doping
is reduced, one could hope to see signs of a reduction in the rigidity
of the vortex lattice.

\begin{theacknowledgments}
This work was supported by the Natural Science and Engineering
Research Council of Canada.

\end{theacknowledgments}




\bibliographystyle{aipproc}   

\bibliography{mrst-talk}

\IfFileExists{\jobname.bbl}{}
 {\typeout{}
  \typeout{******************************************}
  \typeout{** Please run "bibtex \jobname" to obtain}
  \typeout{** the bibliography and then re-run LaTeX}
  \typeout{** twice to fix the references!}
  \typeout{******************************************}
  \typeout{}
 }

\end{document}